# All-optical *in vivo* photoacoustic tomography by adaptive multilayer temporal backpropagation

Taeil Yoon[1,2], Hakseok Ko[2], Jeongmyo Im[3], Euiheon Chung[3], Wonshik Choi[2,*] and Byeong Ha Lee[1,*]

[1]*School of Electrical Engineering and Computer Science, Gwangju Institute of Science and Technology (GIST), Gwangju 61005, Korea*
[2]*Department of Physics, Korea University, Seoul 02841, Korea*
[3]*Department of Biomedical Science and Engineering, Gwangju Institute of Science and Technology (GIST), Gwangju 61005, Korea*
[*]*wonshik@korea.ac.kr, leebh@gist.ac.kr*

**Abstract**

Photoacoustic tomography (PAT) offers high optical contrast with acoustic imaging depth, making it essential for biomedical applications. While many all-optical systems have been developed to address limitations of ultrasound transducers, such as limited spatial sampling and optical path obstructions, measuring surface displacements on rough and dynamic tissues remains challenging. Existing methods often lack sensitivity for *in vivo* imaging or are complex and time-consuming. Here, we present an all-optical PAT system that enables fast, high-resolution volumetric imaging in live tissues. Using full-field holographic microscopy combined with a soft cover layer and coherent averaging, the system maps surface displacements over a 10×10 mm² area with 0.5 nm sensitivity in 1 second. A temporal backpropagation algorithm reconstructs 3D images from a single pressure map, allowing rapid, depth-selective imaging. With adaptive multilayer backpropagation, the system achieves imaging depths of up to 5 mm, with lateral and axial resolutions of 158 µm and 92 µm, as demonstrated through *in vivo* imaging of mouse vasculature.

**Introduction**

Photoacoustic tomography (PAT) offers volumetric imaging with high optical contrast and acoustic imaging depth, making it a vital tool for biomedical applications and industrial inspections (*1*). Its principle involves irradiating short laser pulses onto light-absorbing objects embedded in a scattering medium, where short-duration acoustic waves, called photoacoustic waves, are induced by photo-thermal expansion at the absorption sites. By capturing the 2D pressure profiles of these acoustic waves at the surface of a specimen over time, the 3D distribution of the objects can be visualized (*2*, *3*). Essentially, PAT can selectively visualize the objects having high optical absorption at the excitation laser wavelength (*4*, *5*).

Since the advent of PAT, ultrasound transducers have been widely used to detect the acoustic pressures induced by photoacoustic waves (*6*). When using a single-element transducer, scanning over an area is required, making tomographic reconstruction time-consuming. For efficient volumetric imaging, various forms of transducer arrays have been employed (*7–11*). However, ultrasound transducers face intrinsic limitations, including low spatial sampling bandwidth due to the finite sensor size, which is often larger than the acoustic wavelength, a limited number of sensor elements, and difficulty in securing an optical beam path for excitation laser pulses.

Various optical detection systems have been proposed to address the limitations associated with conventional ultrasound transducers (*12–16*). These methods detect the surface displacements induced by photoacoustic waves, allowing for optical-resolution spatial sampling of acoustic pressure and providing optical windows for the excitation laser pulses. For example, Jathoul *et al.* used a Fabry-Perot polymer film, composed of partially reflective glasses and a polymer spacer, to enhance pressure sensitivity beyond that of transducers (*16*). However, this method requires lateral scanning of the probe laser focus across the film, resulting in a few minutes of measurement time. To enhance surface profiling speed, wide-field detection methods have been developed, enabling the acquisition of surface displacement maps over a large area at once (*17–21*). Nuster *et al.* utilized a phase contrast imaging technique to measure the acoustic pressure integrated along the probe beam direction (*17*, *18*). However, this method requires transmission-based measurements, making it difficult to apply to *in vivo* imaging, where reflection measurements are essential. Buj *et al.* applied an off-axis holography-based optical interferometric system, operating in a pump-and-probe mode, to capture 2D snapshot images of propagating photoacoustic waves (*21*). While the study demonstrated the feasibility of imaging a silicon tube phantom placed under porcine skin, imaging in live tissues has remained out of reach mainly because the system's sensitivity was insufficient to capture photoacoustic waves on a rough and dynamic tissue surface. Additionally, the 3D image reconstruction was performed in the spatio-temporal frequency domain, requiring the recording of a complete set of surface displacement maps over a time span of the photoacoustic wave propagation. Thus, most of the all-optical PAT systems developed so far either lack the sensitivity needed for *in vivo* imaging or are too complex and time-consuming.

Here, we present an all-optical photoacoustic tomography that provides high-speed volumetric imaging with high sensitivity for live tissue imaging. The detection system is based on full-field holographic microscopy, and we use a soft and transparent cover layer on the tissue surface to provide an optically smooth surface for displacement measurements. Coherent averaging of multiple holograms is employed to further enhance the signal-to-noise ratio (SNR) of the phase detection. As a result, our system achieves 0.5 nm displacement sensitivity at each 31 µm optical resolution spot, offering pressure sensitivity comparable to an ultrasound transducer of the same size, while profiling the surface displacements over a 10×10 mm² area in just 1 second. Furthermore, we developed a temporal backpropagation algorithm that can reconstruct a volumetric image from a single surface pressure map, generated from two consecutive displacement maps, enabling rapid and depth-selective volumetric imaging. This algorithm incorporates adaptive multilayer backpropagation, adjusting acoustic velocities *in situ* across different media and correcting angle-dependent attenuation caused by impedance mismatches between the cover layer and the tissue medium. Our system achieves imaging depths of up to 5 mm, with 158 µm lateral and 92 µm axial resolution, demonstrated through *in vivo* volumetric imaging of mouse hindlimb vasculature and chicken embryo blood vessels.

## Results

### All-optical full-field imaging of photoacoustic wave propagation

Optical imaging of photoacoustic waves in living specimens is challenging due to the extremely small surface displacements caused by acoustic pressure, often just a few nanometers or less (*22*). These minute deformations are easily masked by the inherent surface roughness of biological tissues, which can range from tens to hundreds of micrometers. Additionally, the continuous movement of living tissues further complicates the wave detection. To address these issues, we apply an optically transparent and smooth cover layer on the tissue surface (Fig. 1A). In our experiments, we used polydimethylsiloxane (PDMS), a biocompatible material commonly used in wearable technology, as the cover layer (*23*, *24*). This cover layer allows clear optical access to the excitation light while significantly improving the measurement sensitivity of surface displacements. However, the acoustic impedance mismatch between the cover layer and the biological tissue complicates the reconstruction process. As we shall explain in the following, we developed a specialized multilayer backpropagation algorithm to reconstruct the initial photoacoustic pressure.

Our experimental system employed a pump-probe configuration to capture temporal snapshots of the surface displacements induced by photoacoustic wave propagation. An excitation laser pulse (wavelength: 532 nm, pulse width: 8 ns, repetition rate: 20 Hz) is irradiated onto objects embedded in a host medium (Fig. 1A). The generated photoacoustic wave deforms the specimen surface at depth $z = 0$ (Fig. 1B) and further propagates to the top surface of the cover layer at $z = -d$, where it induces surface displacements. A probe laser (wavelength: 905 nm, pulse width: 20 ns, repetition rate: 40 Hz), with a different wavelength from the excitation laser, is used to measure these surface displacements. An off-axis holography system records the 2D interferogram between the probe pulse reflected from the surface of the cover layer and the reference pulse (see Supplementary Information section 1). By applying the Hilbert transform to the interferogram, we obtain the complex-field map—i.e., amplitude and phase maps—of the probe pulses. From the phase map $\phi(x, y, z = -d, t)$, the surface displacement map $\xi(x, y, z = -d, t)$ can be calculated using the relation $\xi = (\lambda/4\pi)\phi$, where $\lambda$ is the wavelength of the probe pulse.

The excitation beam had a 5 mm diameter and a fluence of 15 mJ/cm², adhering to the American National Standards Institute (ANSI) laser safety standards. The probe beam, with a diameter of 16 mm, covered a field of view of 10×10 mm² in the holographic imaging system. The optical resolution was set to 31 μm, approximately six times smaller than the typical photoacoustic wavelength (~200 μm), providing sufficient spatial sampling bandwidth. This setup achieved spatial sampling with over 104,000 effective elements in each single-shot recording, far exceeding the 256 to 1024 elements typically used in transducer arrays (*10*).

To capture the time evolution of the photoacoustic waves, we recorded the surface displacement maps $\xi(x, y, z = -d, t)$ in a time sequence by adjusting the time delay $t$ between the excitation and probe pulses (Fig. 1C). With a 50 ns interval over a 5 μs time span, we could sample acoustic frequencies up to 10 MHz and reconstruct the photoacoustic image to a depth of up to 5 mm. Although we measured the surface displacement induced by the acoustic wave, it is the acoustic pressure that satisfies the wave equation. Therefore, we obtained the pressure map (Fig. 1D) by taking the time derivative of the displacement as:

$$p(x, y, z = -d, t) = \frac{Z}{2} \frac{\partial}{\partial t} \xi(x, y, z = -d, t). \tag{1}$$

Here, $Z$ is the acoustic impedance of the medium (*12, 25*). As we shall explain below, we obtained the initial volumetric pressure image by backpropagating these pressure waves in time.

To minimize the effect of dynamic surface fluctuations, we employed a differential measurement technique, normalizing the complex-field maps recorded with and without the excitation pulse. This method allowed us to isolate the surface displacements induced solely by the photoacoustic wave (Fig. 1C). Since the probe pulse repetition rate is twice that of the excitation pulse, a single differential measurement could be made per excitation laser pulse, enabling the acquisition of one wide-field displacement map every 0.05 seconds. However, during *in vivo* experiments, this approach was often insufficient for detecting weak photoacoustic signals. To enhance phase detection sensitivity, we used coherent averaging of the complex-field maps, improving the SNR in proportion to the number of averaged maps. By typically averaging 20 displacement maps over 1 second, we achieved a displacement sensitivity of 0.5 nm, corresponding to a pressure sensitivity of approximately 15 kPa.

This sensitivity level is comparable to that of the ultrasound transducer having a size similar to the optical resolution.

**Single-sheet multilayer backpropagation algorithm**

The goal of photoacoustic imaging is to reconstruct the 3D spatial distribution of acoustic pressure at the time of excitation ($t = 0$), known as the initial volumetric pressure image, $p(\mathbf{r}, t = 0)$, which reflects the positions and densities of molecules absorbing the optical excitation pulses. In the case of mapping the pressure on a surface, the initial pressure distribution can be reconstructed by backpropagating the acoustic waves using angular spectrum analysis (*26*, *27*). In this approach, the spatiotemporal spectrum of the acoustic waves is obtained by applying a Fourier transform to the surface pressure maps, $p(x, y, z = -d, t)$, in both lateral space and time. Each spectral component is then backpropagated according to its temporal frequency. The initial volumetric pressure image is reconstructed by summing the backpropagated waves across the entire spatiotemporal spectrum (*26–28*). While this reconstruction process is simple and straightforward, it cannot begin until all time-sequential measurements are complete, limiting its application for imaging dynamic samples.

To overcome this limitation and enable rapid imaging, we developed a reconstruction algorithm termed the single-sheet backpropagation (SSB) method, which backpropagates the 2D surface pressure map taken at each time delay into the 3D volume. For each surface pressure map $p(x, y, z = -d, t')$ taken at a specific time $t'$, the lateral spatial spectrum is obtained by applying the Fourier transform with respect to lateral coordinates:

$$P(\mathbf{q}, z = -d, t') = \iint p(x, y, z = -d, t') e^{-ik_x x} e^{-ik_y y} dx dy. \qquad (2)$$

This spatial spectrum represents the amplitude of the acoustic wave with a lateral wave vector $\mathbf{q} = (k_x, k_y)$. The wave is first backpropagated to the bottom surface of the cover layer at $z = 0$. Since the photoacoustic wave is a pulse, a wave component having a specific lateral wave vector is composed of many plane waves with different temporal frequencies $\omega$ and, thus, different longitudinal wave vectors $\beta$, called the propagation constant (Fig. 2A). Here, $\omega$ and $\beta$ are related by the dispersion relation via the acoustic velocity $v_c$ in the cover layer:

$$\omega = v_c\sqrt{|\mathbf{q}|^2 + \beta^2}. \tag{3}$$

Then, each plane wave constituting the same lateral wave vector $\mathbf{q}$ can be backpropagated spatially to the depth $z = 0$ and temporally to the time $t$:

$$P(\mathbf{q}, z = 0, t; t') = \int A(\omega) P(\mathbf{q}, z = -d, t') e^{-i\beta d} e^{i\omega(t'-t)} d\beta. \tag{4}$$

Here, $A(\omega)$ is the magnitude of the temporal spectrum of the photoacoustic wave, calibrated from a prior measurement (Figs. 2B and C). Since the complex amplitude of each acoustic plane wave in the right-hand side of Eq. (4) evolves over time with its temporal frequency $\omega$, we obtain the wave at a previous time $t$ by backpropagating it for a time duration of $(t' - t)$. The main assumption in Eq. (4) is that the temporal response of the photoacoustic wave remains largely the same regardless of the spatial location of the object and the time delay. This allows the use of the calibrated temporal response of the system.

To find the initial volumetric pressure image within the medium, the acoustic wave of each lateral wave vector should be propagated further back to a depth $z$ within the medium:

$$P(\mathbf{q}, z, t; t') = \int A(\omega) P(\mathbf{q}, z = -d, t') e^{-i\beta d} e^{-i\beta_m z} e^{i\omega(t'-t)} d\beta. \tag{5}$$

Here, $\beta_m$ is the propagation constant in the host medium. Since the temporal frequency of a plane wave does not change with propagation, and the lateral spatial frequency remains constant due to continuity at the boundary, $\beta_m$ is determined by the acoustic velocity $v_m$ in the medium and $\beta$:

$$\beta_m = \sqrt{\left(\frac{v_c^2 - v_m^2}{v_m^2}\right)|\mathbf{q}|^2 + \left(\frac{v_c}{v_m}\right)^2 \beta^2}. \tag{6}$$

Finally, the initial volumetric pressure image is calculated by taking the inverse Fourier transform of the lateral spectrum obtained by backpropagating in time to the moment of $t = 0$ in Eq. (5):

$$p(\mathbf{r}, t = 0; t') = \iint P(\mathbf{q}, z, t = 0; t') e^{ik_x x} e^{ik_y y} dk_x dk_y. \tag{7}$$

The SSB algorithm enables the reconstruction with a single surface pressure map taken at a specific time $t'$. This relieves the need for taking a full set of pressure maps in a time series, making it more suitable for imaging dynamic biological processes. However, the reconstruction is limited in the depth range because only the waves captured at a specific time can be involved in the reconstruction process. Waves that have not yet reached the surface or have already passed away at the time $t'$ cannot be used for the reconstruction. We can increase the imaging volume by the coherent superposition of multiple images backpropagated from surface pressure maps taken at different time delays, which also leads to the increased SNR (see Supplementary Movie S1).

Figure 2 illustrates the concept of the proposed SSB method. Figure 2A shows the backpropagation of each lateral wave vector **q** constituting the acoustic wave captured at a specific time $t'$. The images in each row depict the backpropagation of waves having the same **q**, but with different $\omega$ and $\beta$. In the figure, $\omega_{ij}$ represents the temporal angular frequency of the plane wave with $i$-th **q** and $j$-th $\beta$. The initial pressure distribution is reconstructed by reversing the time evolution of all these plane waves from the captured moment back to the time of excitation. During time reversal, the phase of each plane wave evolves according to its temporal frequency. The spectrum $A(\omega)$ of the generated photoacoustic waves was estimated by sequentially measuring a set of photoacoustic waves emanating from a line-shaped object. Figure 2B shows the temporal spectra calculated along the object's length (see Supplementary Information section 2). The spectrum in Fig. 2C (blue solid curve) was obtained by averaging along the vertical direction, and the red dotted curve shows the curve fitting.

Figure 2D shows normalized maximum intensity projection images of a knot-shaped object, reconstructed using 1, 3, and 20 surface pressure maps acquired at 50 ns intervals. The 3D view of the object, reconstructed with 20 maps, is shown in Fig. 2E. Even with a single pressure map, the target object could be reconstructed, though the SNR was improved with additional maps. To validate the tomographic reconstruction capability, a phantom consisting of two line-shaped objects with a 1 mm depth separation was imaged. The first two sub-images in Fig. 2F, reconstructed from two single surface pressure maps taken at different times, separately visualized the upper and lower objects. The third image, reconstructed from 40 maps over a 2 μs span, shows both objects. Its 3D view is presented in Fig. 2G, confirming that the reconstruction depth can be expanded by increasing the number of pressure maps used.

**PAT in scattering media by coherent averaging and adaptive backpropagation**

The proposed method was experimentally verified by imaging the objects embedded within an optically thick scattering medium. Several pieces of black polyethylene terephthalate (PET) fibers, each with a diameter of 200 μm, were arranged in the form of 'AOL' alphabet characters (top of Fig. 3A). The PDMS medium was prepared by mixing silicone and curing agent in a 10:1 ratio and adding 0.3% by weight of TiO$_2$ particles. Due to multiple light scattering in the medium, the embedded PET fibers were not optically visible (bottom of Fig. 3A). Since the size of the 'AOL' phantom was larger than the size of the excitation beam, four sub-regions were individually imaged and stitched together after reconstruction. The total imaging area is indicated by the green box in Fig. 3A.

Figure 3B shows a few representative time-evolving displacement maps. Photoacoustic waves with nanometer-scale displacement amplitudes traversing the surface were identified; however, there was substantial noise. To reduce the noise, 50 complex-field maps were captured at each time delay and coherently averaged after compensating for the overall phase drift during multiple measurements (see Supplementary Information section 3). This increased the SNR from 6 to 12 dB (Fig. 3C). The achieved phase sensitivity was 6.9 mrad, corresponding to a displacement amplitude of 0.5 nm. The SNR increased with the number of images used for averaging but saturated at around 20 images. Therefore, 20 displacement maps were recorded at each time delay in all the photoacoustic imaging results presented in this study.

In backpropagation, the proper setting of the acoustic velocity is critical for achieving resolution and contrast in reconstruction (*29–32*). The velocity depends on the type of medium and the frequency $\omega$ through the dispersion relation (*33*). For *in vivo* experiments, determining the acoustic velocity is challenging because the medium is a heterogeneous mixture of various biomolecules. To address this, we developed an adaptive backpropagation algorithm to find the acoustic velocities *in situ*. Specifically, we expanded the velocity in the medium to the first order of $\omega$:

$$v(\omega) = v_0 + \alpha\omega. \tag{9}$$

We then find the coefficients in such a way as to maximize the image sharpness metric $S$ defined as (*34–36*):

$$S = \sum_{x,y,z}\bigl(p_0(x, y, z, t = 0; v_0, \alpha)\bigr)^4. \tag{10}$$

The choice of the metric is somewhat flexible, depending on factors such as the structure of the object and the quality of experimental data.

The reconstructed image was somewhat blurred (Fig. 3D) when we used the acoustic velocity of 1076 m/s, an estimated value from the literature for PDMS with a 10:1 ratio (*37*). After applying the adaptive backpropagation algorithm, the acoustic velocity was determined to be 1045 m/s. The reconstructed image shown in Fig. 3E visualizes the fine structure, particularly near the knot part indicated by a white arrow. The measured width of the 200 μm diameter PET fiber was 262 μm with a velocity of 1076 m/s, but it was reduced to 204 μm with 1045 m/s. It should be noted that, due to the round shape of the fiber, only a small portion of its cross-section contributed to the photoacoustic wave generation.

The velocity estimated by the adaptive backpropagation algorithm was independently verified using a simple trigonometric method. Specifically, by measuring the distance between counter-propagating photoacoustic waves emanating from a line-like object, the velocity was calculated to be 1044 m/s, which is close to the *in situ* optimized velocity (see Supplementary Information section 4). We also found that $v_0$ played a more important role than $\alpha$, suggesting that the effect of dispersion was not significant. When applying the adaptive backpropagation to a multilayer sample with a cover layer, the algorithm has been designed to determine the optimal values of $v_0$ and $\alpha$ in each medium.

**Multilayer acoustic backpropagation for cover-layered phantoms**

As mentioned earlier, we introduced a cover layer to smooth out the rough and dynamically fluctuating surface of the specimen, enabling precise optical mapping of photoacoustic waves. Among various materials considered—such as acrylic, glass, and quartz—we selected a PDMS plate for the cover layer due to its biocompatibility, mechanical flexibility (*23*, *24*), and acoustic impedance matching. In our experiment, we placed two pencil leads with a diameter of 300 μm in a cross pattern inside a petri dish filled with deionized water (Fig. 4A). The dynamic fluctuations of the water surface made photoacoustic wave mapping challenging, but by placing a 3.44-mm-thick PDMS block on top of the water-filled dish, we were able to stabilize the surface (Fig. 4B).

The photoacoustic waves were recorded at the top surface of the PDMS cover layer rather than the bottom. This preference is due to the higher contrast in the holograms at the top surface (Fig. 4B), which translates to enhanced phase detection sensitivity. The optical reflectivity at the top surface (air vs. PDMS) is 48 times larger than that at the bottom surface (PDMS vs. water), making the top surface more suitable for measurement (Fig. 4C). Additionally, the irregular surface of the specimen at the bottom negatively affected the optical reflectivity distribution. Despite all these benefits, measurements at the top surface required multilayer backpropagation for accurate image reconstruction, as discussed earlier.

Figure 4E presents a 3D image reconstructed using our adaptive multilayer backpropagation algorithm. The acoustic velocities were optimized based on the sharpness metric in Figure 4D, where $v_c$=1020 m/s in the cover layer and $v_m$=1541 m/s in the water yielded the highest sharpness. The reconstructed 3D image clearly visualized the two pencil leads; however, slight blurring remained even after correcting for the acoustic velocities. This was attributed to the acoustic Fresnel reflections at the cover-specimen boundary. Due to the impedance mismatch, waves with higher propagation angles experienced more pronounced Fresnel reflection, resulting in increased attenuation. To address this, a correction factor $B(\mathbf{k})$ was applied to compensate for the angle-dependent attenuation depending on the propagation angle. As shown in Figure 4F, this adjustment significantly improved the image sharpness from 349 μm to 232 μm.

### *In vivo* vasculature imaging of mouse hindlimb and chicken embryo

To validate the utility of the proposed all-optical photoacoustic imaging method, we have tried to image the vasculature of a living mouse. We prepared a 5-week-old nude mouse (Fig. 5a) and placed a 2.28-mm-thick PDMS block on its hindlimb. The measurement was made at the top surface of the PDMS block, and the acoustic velocities of 1030 m/s was used for the cover layer and 1540 m/s for the mouse. To expand the field of view, the sample stage was laterally translated to three adjacent regions.

Figure 5d presents the reconstructed volumetric image of the vasculature. We successfully visualized the main blood vessels distributed at depths ranging from 1.2 to 2 mm beneath the skin. The figure distinctly shows two major blood vessels, running parallel to each other, along with additional vascular features in the upper portion of the image. To validate the imaging results, we dissected the skin of the mouse right after the photoacoustic imaging. Figure 5b reveals that fat tissues were situated beneath the skin layer, and they overlaid the upper part of the main blood vessels. After removing the fat tissues, we could clearly observe the specific blood vessels designated as area 1 in Fig. 5c. In imaging area 2, we could see that the blood vessels were composed of femoral artery and vein. The photoacoustic image in Fig. 5d was in excellent agreement with the anatomy in Figs. 5b and 5c. Interestingly, the photoacoustic image could visualize the vasculature even below the fat tissues, which were optically so thick that we could not see any blood vessels with photography. These results validate the accuracy and utility of our approach for *in vivo* applications.

As another application, we have performed *in vivo* blood vessel imaging of a chicken embryo. Figure 5e shows a photograph of an 11-day-old chicken embryo. Zooming into the region of interest, indicated by the black circle in the left-side image of Fig. 5e, reveals well-developed blood vessels on the CAM. Measuring the surface displacement in this environment was particularly challenging due to its pulpy surface, which was susceptible to external vibrations or biological factors like heartbeats. The issue of pulpy imaging surface could be resolved in this study by placing a 2.42-mm-thick PDMS cover layer on top of the CAM. A gap of approximately 1~2 mm between the cover layer and the CAM surface was filled with normal saline to prevent the blood vessels from rupturing due to osmotic pressure. We adaptively identified the acoustic velocities of 1035 m/s in the cover layer and 1560 m/s in the normal saline. The reconstructed image of Fig. 5f shows the complex vasculature of the CAM with high contrast, which is well matched with the photographic evidence in Fig. 5e.

**Discussion**

In this study, we propose a photoacoustic tomography technique capable of measuring photoacoustic waves even on a rough and dynamic tissue surface with high sensitivity and speed. The proposed method achieved surface profiling across a 10×10 mm² field of view with a spatial resolution of 31 µm and 0.5 nm displacement sensitivity, enabling surface pressure mapping with sensitivity comparable to that of an ultrasound transducer of a similar size with the optical resolution. This system provides approximately $10^5$ sampling elements, two orders of magnitude more than conventional ultrasound transducer arrays. Moreover, the system captures one full-field image, including coherent averaging, in just one second, offering a sampling rate of $10^5$ per second—over ten times faster than beam-scanning methods, assuming the same excitation laser repetition rate (*16*).

Additionally, we introduce a new 3D reconstruction method termed single-sheet backpropagation. Traditional backpropagation requires measuring the full-time span of the photoacoustic wave propagation to initiate reconstruction. In contrast, our method allows image reconstruction from each temporal snapshot. While the imaging depth range depends on the time span, our approach is better suited for imaging dynamic samples when investigating specific depth ranges. We further developed methods to identify acoustic velocities within different media during post-processing and to correct for angle-dependent attenuation caused by impedance mismatch between layers, optimizing reconstruction contrast and resolution.

One of the key ideas in this study is the application of an optically smooth and soft PDMS cover layer to flatten the rough tissue surface. While we also experimented with acrylic, glass, and quartz plates, the PDMS cover layer produced the best reconstruction images. This is because the other plates had higher acoustic impedance than the tissues, causing total internal reflection to the high-angle photoacoustic waves and limiting their propagation into the cover layer. Additionally, the higher acoustic velocity in these plates led to repeated reflections of the photoacoustic waves within the cover layer, producing duplicated and overlapped images.

Looking ahead, several improvements and applications could further enhance the proposed photoacoustic tomography technique. First, integrating a high-repetition-rate laser would enable real-time volumetric imaging, significantly expanding the technique's utility for *in vivo* applications and dynamic tissue monitoring. Moreover, advancements in adaptive algorithms for reconstructing images in more complex and heterogeneous media could allow for higher accuracy and deeper imaging depth in challenging environments, such as through-skull imaging. This system can also be optimized for broader biomedical applications, such as functional imaging to assess tissue oxygenation, molecular imaging, or tumor detection with high optical contrast. Additionally, the technique has potential applications beyond biomedicine, such as in industrial non-destructive testing, where it could be used for detecting flaws in semiconductor wafers, inspecting civil structures, or remotely monitoring the integrity of high-voltage electrical equipment (*38*). With further refinement and expansion, the proposed PAT system could become a versatile tool across various fields, offering high-resolution imaging with unprecedented speed and sensitivity.

**Methods**

**Experimental setup**

A Nd:YAG laser (Quantel, Ultra 50, France) with a wavelength of 532 nm and a pulse width of 8 ns was used for photoacoustic excitation. Köhler illumination was employed to uniformly irradiate the excitation beam onto the target objects embedded in the specimen medium (*39*, *40*). To probe the surface displacement, a fiber-pigtailed pulsed diode laser (PLD; LASER COMPONENTS, 905D2S3J09FP-40, Germany), with a wavelength of 905 nm and a pulse width of 20 ns, was used. An off-axis holography system was implemented by introducing a tilted reference mirror. Each interferogram, called a digital hologram, was captured by a sCMOS camera (scientific complementary metal-oxide-semiconductor camera; Andor, Zyla 4.2, Japan) with 1024×1024 elements and a pixel size of 6.5 μm. The captured hologram was Fourier transformed to obtain its spatial frequency spectrum, and one of the twin cross-correlation parts of the spectrum was spatially filtered and re-centered. Finally, the complex field recorded in the hologram was retrieved by taking the inverse Fourier transform. By extracting the phase part of the complex field and multiplying it by the optical wavelength of the probe beam, the 2D surface profile of the specimen was obtained. The net surface displacement map, induced solely by the photoacoustic waves, was determined by taking the difference between the maps captured with and without the excitation beam. The phase change of $2\pi$ radians was treated as the displacement corresponding to half the wavelength of the probe beam, accounting for its round trip.

The PLD was operated at 40 Hz to sequentially capture holograms with and without excitation. The sCMOS camera was synchronized to the trigger of the Nd:YAG laser, which was operating at 20 Hz. The time delay between the excitation and probe pulses was adjusted with 50 ns intervals using a programmable pulse generator. While the exact number of time delays depends on the imaging depth, in the experiment, there were approximately 20 or more, corresponding to a 5 mm imaging range along the depth direction in biological tissue with a typical acoustic velocity of 1540 m/s (*41*).

**Animal preparation**

**Nude mouse:** A 5-week-old nude mouse was prepared for small animal vascular imaging. To minimize mouse movement during the experiment, the mouse was anesthetized using an intraperitoneal injection of Zoletil/Xylazine at a dose of 60/10 mg/kg body weight. Then the animal was placed on a heating plate for imaging. After the experiment, the mouse was returned to its cage and placed on another heating pad to aid recovery. A light-dark cycle of 12 hours was established, and conditions were maintained to ensure a stable body temperature. The animal experiments followed the protocol approved by the Institutional Animal Care and Use Committee (IACUC) of Gwangju Institute of Science and Technology (GIST).

**Chicken embryo:** To image the blood vessels of the chicken embryo's CAM, fertilized eggs were purchased in a market and incubated for 11 days in an incubator. The temperature and humidity within the incubator were maintained at 37°C and 60%, respectively. Following incubation, the chicken embryo exhibited a heart rate of 126 beats per minute at room temperature (25°C). To minimize the movement caused by heartbeats, the embryos were placed in a refrigerator of 5°C for 30 minutes. This decreased the heart rate down to 60 beats and slowed down the body movement. To expose the CAM, a section of the outer shell and the inner membrane of the air cell were removed, and a PDMS cover layer was positioned atop the remaining outer shell. The gap between the outer shell and the cover layer was filled with normal saline to prevent blood vessels from rupturing due to osmotic pressure.

**Acknowledgments**

**Funding:** This work was supported by the Technology Innovation Program (20021979, Development of health monitoring system for gas-insulated switchgears by using optical sensing modalities) funded by the Ministry of Trade, Industry and Energy (MOTIE, Republic of Korea) and GIST Research Project grant funded by GIST in 2024. H.K. and W.C. are supported by the Institute for Basic Science (IBS-R023-D1) and the National Research Foundation of Korea (NRF) grant funded by the Korea government (MSIT) (No. RS-2024-00442818). J.I. and E.C. are supported by the NRF grant funded by the Korea government (MSIT) (No. RS-2023-00302281).

**Author contributions:** T.Y., H.K., W.C., and B.L. conceived the experiment, and T.Y. performed experiments and analyzed the experimental data. J.I. and E.C. prepared animal experiments and investigated vascular architecture. T.Y., W.C. and B.L. developed the theoretical framework and prepared the manuscript. B.L. devised the concept of single-sheet backpropagation (SSB) and T.Y. confirmed it experimentally. All authors contributed to finalizing the manuscript.

**Competing interests:** The authors declare that they have no competing interests.

**Data and materials availability:** All data needed to evaluate the conclusions in the paper are present in the paper and/or the Supplementary Materials.

**Figures**

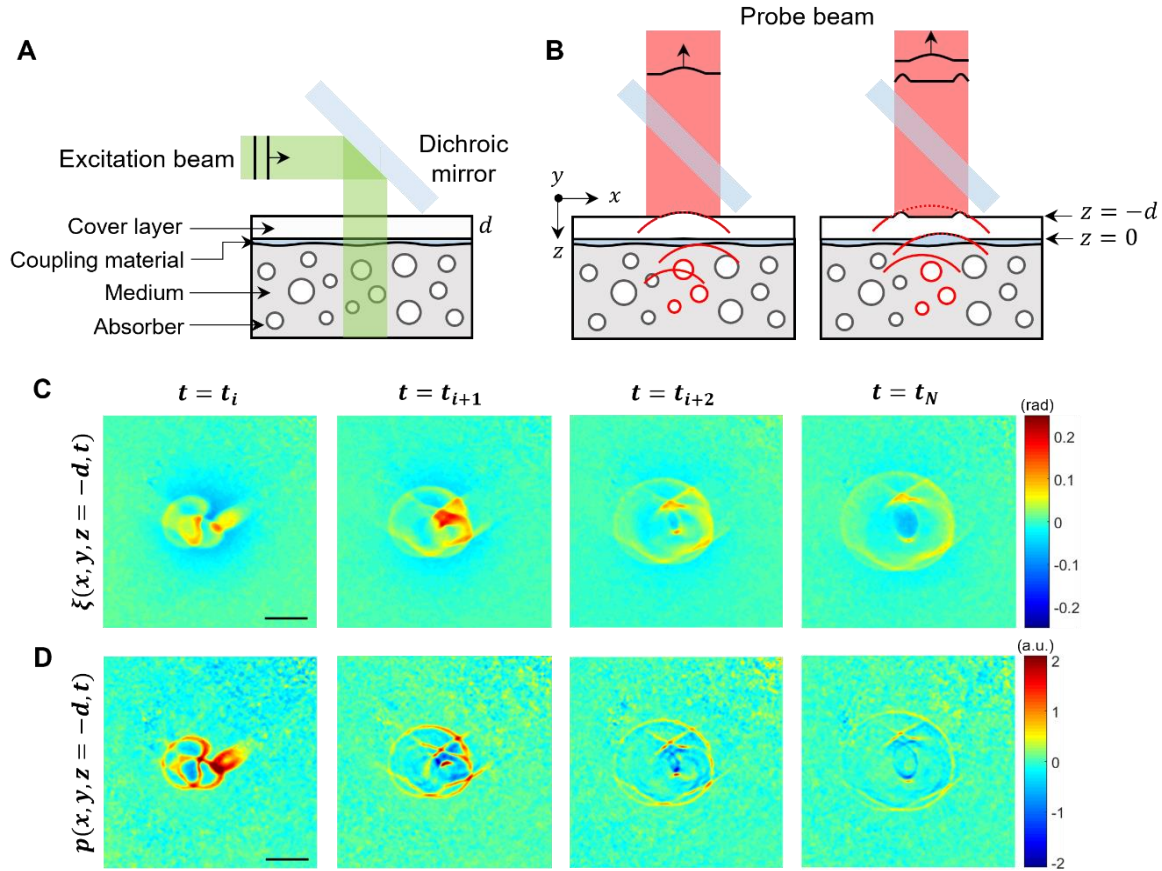

**Fig. 1. Schematic of full-field optical interferometric imaging of photoacoustic waves at the specimen surface.** (**A**) When an excitation pulse with a beam size of several millimeters is irradiated to the specimen through a dichroic mirror, each absorber inside the specimen absorbs light energy and generates a pulse of acoustic wave. A soft cover layer is placed on top to smoothen the rough sample surface. (**B**) The photoacoustic wave propagates toward the surface and induces displacement at the top surface of the cover layer, at $z = -d$. The map of surface displacements is optically measured by taking an off-axis hologram with a pulsed probe beam. (**C**) By adjusting the time delay $t$ between the excitation and the probe pulses, the 2D displacement maps are captured in a time sequence. Color bars: phase in radians. (**D**) By taking the time derivative with the acquired displacement maps, the pressure maps are calculated, which satisfy the wave equation. These maps visualize the expansion or propagation of pressure waves with time. Color bar: scaled pressure. Scale bars in (**C**)-(**D**): 2 mm.

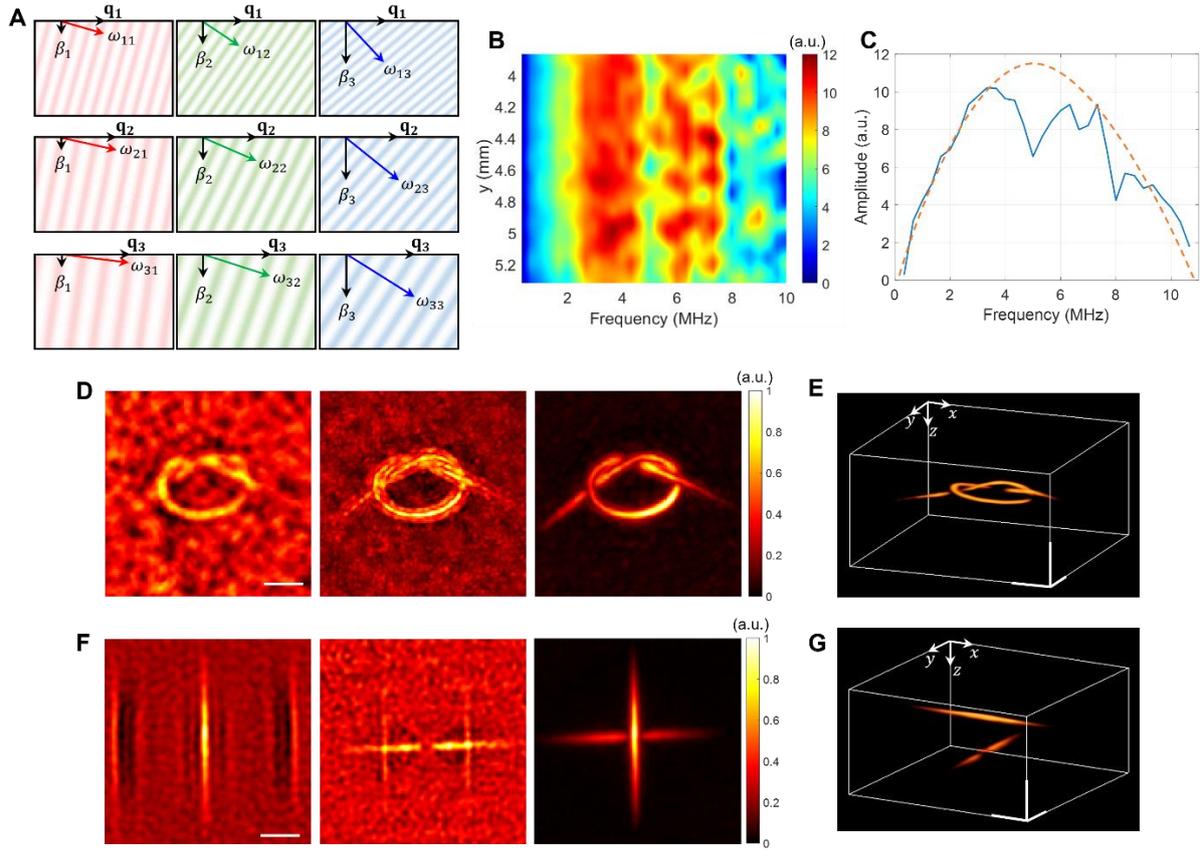

**Fig 2. The single-sheet backpropagation (SSB) method for 3D reconstruction from single surface pressure map.** (**A**) Each row illustrates that the wave having the same lateral spatial frequency **q** can be decomposed with multiple plane waves having different propagation constant $\beta$'s (different columns). The temporal frequency $\omega$ of each plane wave is determined by the combination of **q** and $\beta$ through the dispersion relation of the medium. (**B**) Temporal frequency spectra of surface pressure maps obtained with a line-shaped PET fiber embedded in PDMS. (**C**) Averaged spectrum along the direction ($y$-direction) of the PET fiber fitted with a parabola (red dotted curve). (**D**) Maximum intensity projection images reconstructed by using 1, 3, and 20 surface pressure maps with the proposed SSB method, and (**E**) a 3D view reconstructed using 20 maps. The SNR is enhanced with the number of maps. (**F**) For objects located crosswise with a depth difference of 1 mm, a depth-selective thin-volume image was reconstructed by using a single pressure map (first, second) taken at different times, and the full-volume image by using 40 pressure maps (third). (**G**) The 3D view reconstructed using 40 pressure maps. Scale bars in (**D**)-(**G**): 1 mm.

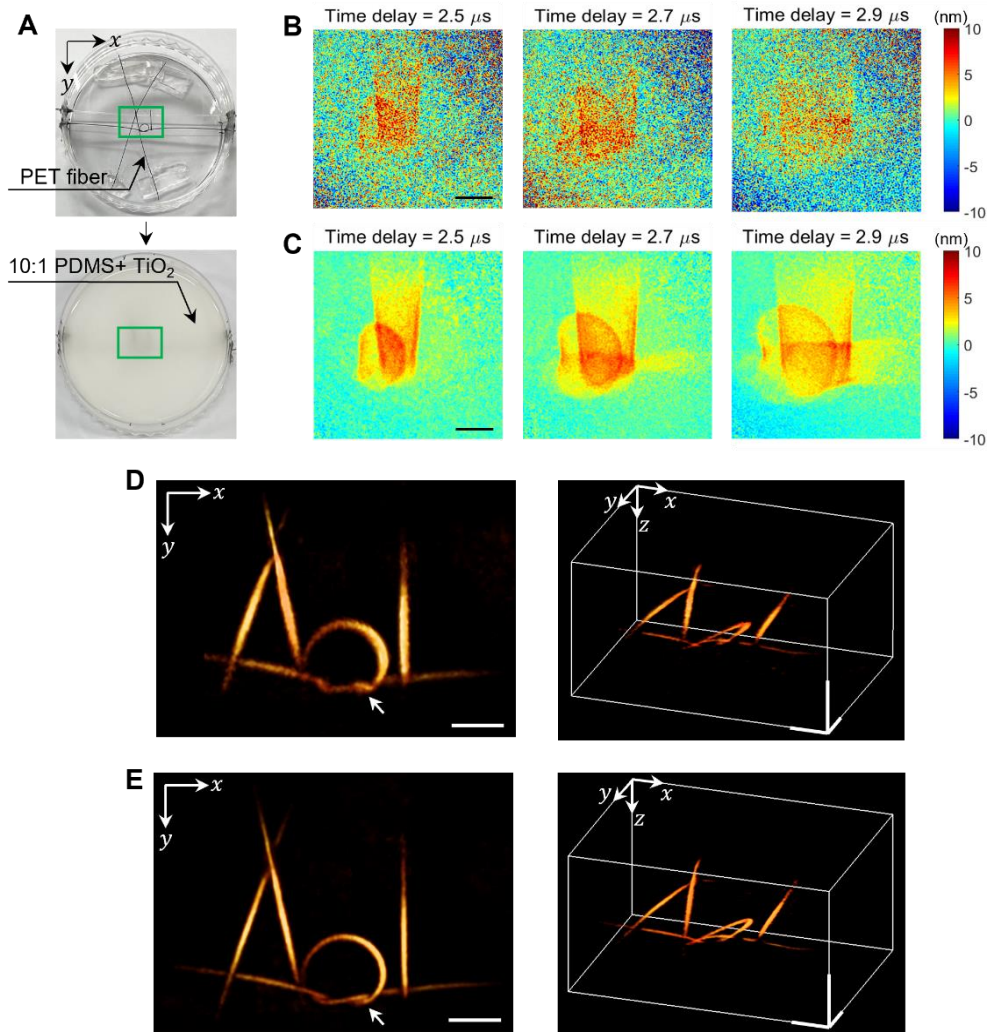

**Fig. 3. Experimental demonstration of coherent averaging and adaptive backpropagation.** (**A**) An 'AOL' shape structure was made with a black PET fiber of 200 μm diameter (top), which was embedded within an optically opaque PDMS made by mixing with 0.3 % weight $TiO_2$ particles (bottom). (**B**) The 2D displacement maps captured at the PDMS surface for different time delays. (**C**) The 20 displacement maps coherently averaged by compensating the overall phase drift. (**D**) The 3D image reconstructed by using the acoustic velocity of 1076 m/s, a literature value for the 10:1 ratio PDMS. (**E**) Reconstructed image by applying adaptive backpropagation algorithm for finding the proper acoustic velocity. At the velocity of 1045 m/s, we could reconstruct the image with optimal contrast and resolution. Scale bars in (**B**)-(**E**): 2 mm.

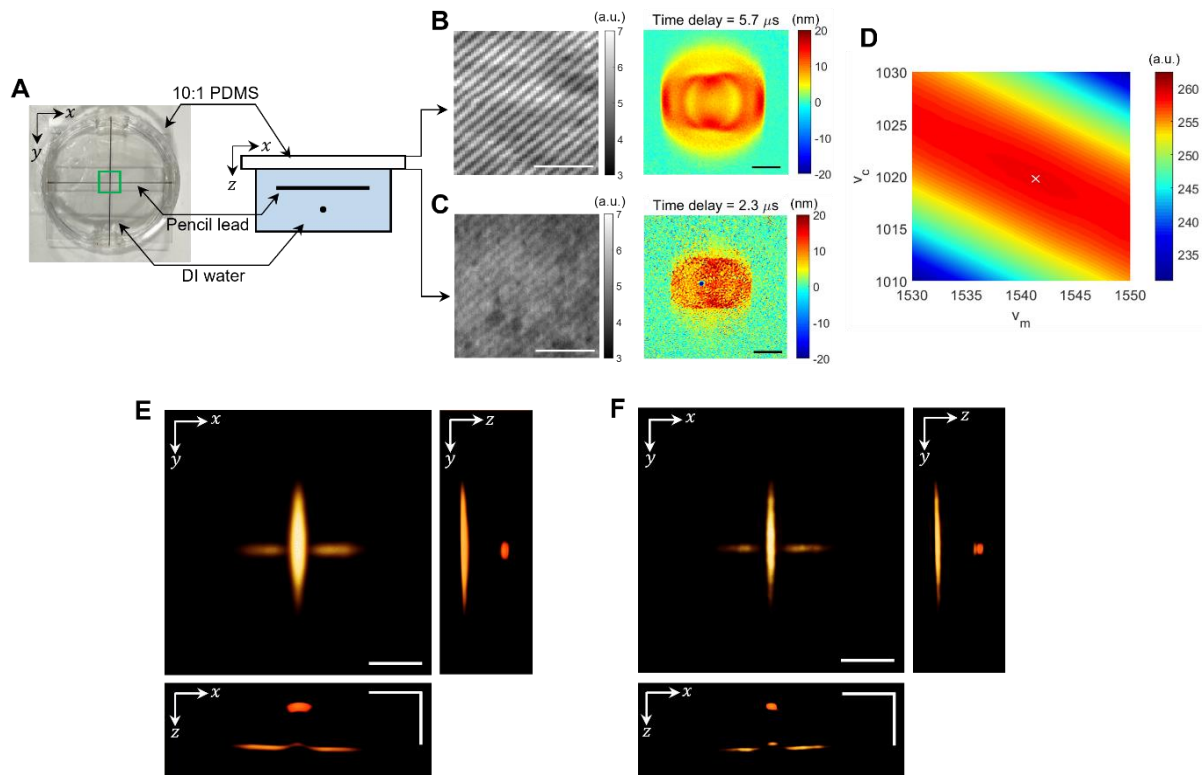

**Fig. 4. Experimental demonstration of the multilayer acoustic backpropagation algorithm.** (**A**) Two 300 μm diameter pencil leads were placed orthogonal to each other at a depth difference of about 1 mm. The phantom was filled with DI water and covered with a 3.44-mm-thick PDMS block. (**B**) The hologram taken at the top surface of the block (left) and the extracted surface displacement map (right). (**C**) Same as (**B**), but taken at the bottom surface of the PDMS block. The white and black scale bars in (**B**)-(**C**): 200 μm and 2 mm, respectively. (**D**) Sharpness metric calculated with various acoustic velocities $v_c$ and $v_m$. (**E**) The 3D image reconstructed without accounting for the propagation angle-dependent attenuation due to acoustic impedance mismatch. (**F**) The same image but reconstructed after compensating the propagation angle-dependent attenuation. Scale bars in (**E**)-(**F**): 2 mm.

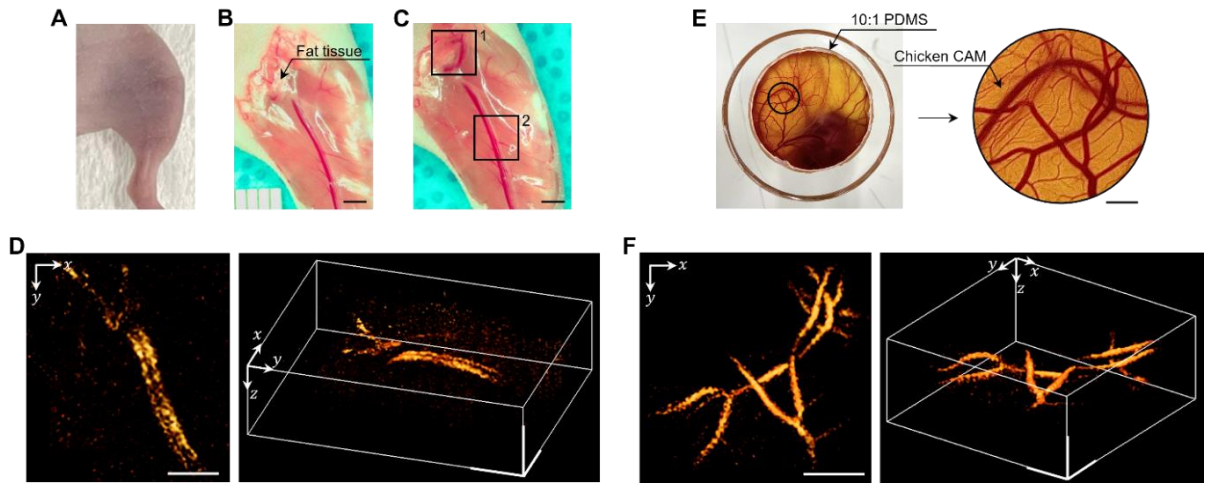

**Fig. 5.** *In vivo* **imaging of vasculatures in mouse hindlimb leg and chicken embryo's CAM.** (**A**) A 5-week-old nude mouse placed in the sample arm of the FF-PAT system for *in vivo* imaging through the intact skin. (**B**) A photograph of the blood vessels in the hindlimb, taken after imaging session, which was used as ground truth. The upper region was covered with fat tissue. (**C**) A photograph taken after further removing the fat tissue revealing vascular structures underneath (area 1). Two vessels of artery and vein are distinctly visible in area 2. (**D**) The 3D image of the mouse hindlimb reconstructed with the multilayer backpropagation method. The blood vessels in areas 1 and 2 in (**C**) are clearly resolved. Scale bars in (**B**)-(**D**): 2 mm. (**E**) The photograph and its enlarged image of the blood vessels formed in the CAM of chicken embryo incubated for 11 days. Scale bar: 1 mm. (**F**) Reconstructed volumetric vasculature image of the CAM. Scale bar: 2 mm.